\documentclass[pra,twocolumn,superscriptaddress,letterpaper]{revtex4}

\usepackage[OT1]{fontenc}
\usepackage[latin1]{inputenc}
\usepackage[english]{babel}
\usepackage{amssymb,amsfonts,amsmath}
\usepackage{graphicx}
\usepackage{xspace}
\usepackage{psfrag}

\usepackage{mathtools}

\DeclarePairedDelimiter\floor{\lfloor}{\rfloor}

\newcommand{\ie}{i.e.\@\xspace}

\newcommand{\eq}[1]{Eq.~\eqref{eq:#1}}

\newcommand{\bm}[1]{\boldsymbol{\mathbf{#1}}}
\newcommand{\ud}{\mathrm{d}}
\newcommand{\bra}{\left\langle}
\newcommand{\ket}{\right\rangle}

\newcommand{\phiout}{\phi_{\mathrm{out}}}
\newcommand{\phioutr}{\phi_{\mathrm{out,R}}}
\newcommand{\phioutt}{\phi_{\mathrm{out,T}}}
\newcommand{\phiin}{\phi_{\mathrm{in}}}
\newcommand{\tout}{t_{\mathrm{out}}}
\newcommand{\tin}{t_{\mathrm{in}}}

\begin{document}

\title{Invariance property of wave scattering through disordered media}

\author{Romain Pierrat}
\affiliation{ESPCI ParisTech, PSL Research University, CNRS, Institut Langevin, 1 rue Jussieu, F-75005 Paris, France}
\author{Philipp Ambichl}
\affiliation{Institute for Theoretical Physics, Vienna University of Technology, Wiedner Hauptstra\ss e 8--10/136, A-1040 Vienna, Austria}
\author{Sylvain Gigan}
\affiliation{Laboratoire Kastler Brossel, Universit\'e Pierre et Marie Curie, Ecole Normale Sup\'erieure, CNRS,
Coll\`ege de France, 24 rue Lhomond, F-75005 Paris, France}
\author{Alexander Haber}
\affiliation{Institute for Theoretical Physics, Vienna University of Technology, Wiedner Hauptstra\ss e 8--10/136, A-1040 Vienna, Austria}
\author{R\'{e}mi Carminati}
\affiliation{ESPCI ParisTech, PSL Research University, CNRS, Institut Langevin, 1 rue Jussieu, F-75005 Paris, France}
\author{Stefan Rotter}
\affiliation{Institute for Theoretical Physics, Vienna University of Technology, Wiedner Hauptstra\ss e 8--10/136, A-1040 Vienna, Austria}

\begin{abstract}
   A fundamental insight in the theory of diffusive random walks is that the mean length of trajectories traversing a
   finite open system is independent of the details of the diffusion process. Instead, the mean trajectory length
   depends only on the system's boundary geometry and is thus unaffected by the value of the mean free path. Here we show that
   this result is rooted on a much deeper level than that of a random walk, which allows us to extend the reach of
   this universal invariance property beyond the diffusion approximation. Specifically, we demonstrate that an
   equivalent invariance relation also holds for the scattering of waves in resonant structures as well as in ballistic,
   chaotic or in
   Anderson localized systems. Our work unifies a number of specific observations made in quite diverse fields of
   science ranging from the movement of ants to nuclear scattering theory. Potential experimental realizations using
   light fields in disordered media are discussed. 
\end{abstract}

\maketitle

In the biological sciences it has been appreciated for some time now that the movement of certain insects
(like ants) on a planar surface can be modeled as a diffusive random walk with a given constant speed $v$
\cite{Turchin1991, Crist1991, Holmes1993}.  Using this connection, Blanco~\emph{et al.}~\cite{BLANCO-2003} proved that
the time that these insects spend on average inside a given domain of area $A$ and with an external boundary $C$ is
independent of the parameters entering the random walk such as, e.g., the transport mean free path (MFP) $\ell^*$.
Specifically, the average time $t$ between the moments when an insect enters the domain and
when it first exits it again, is given by the simple relation $\langle t\rangle=\pi A/(C v)$.
One finds that the mean length $\langle l\rangle$ of the corresponding random walk trajectories inside the domain is also constant,
$\langle l\rangle=\langle t\rangle \, v= \pi A / C$. Similar relations also hold in three dimensions, $\langle
t\rangle=4 V/(\Sigma v)$ and $\langle l\rangle=4 V/\Sigma$, where $V$ is the volume and $\Sigma$ is the external surface
of a given domain. Extensions of this result exist for trajectories beginning inside the domain~\cite{Mazzolo2004}
or for the calculation of averaged residence times inside sub-domains~\cite{Benichou2005}. As a generalization of the
mean-chord-length theorem~\cite{CASE-1967} for straight-line trajectories with an infinite MFP, this fundamental theorem
has numerous applications, e.g., in the context of food foraging~\cite{Campo2011} and for the reaction rates in
chemistry~\cite{Benichou2008}.  

The surprising element of this result can be well appreciated when applied to the physical sciences and, in
particular, to the transport of light or of other types of waves in scattering media. In that context it is well-known
that the relevant observable quantities all do depend on $\ell^*$: In the diffusive regime, the total
transmission of a slab of thickness $L$ scales with $\ell^*/L$ through Ohm's law, and the characteristic dwell time
scales with the so-called Thouless time $L^2/(vi\, \ell^*)$~\cite{Akkermans2007}. When considering coherent wave effects,
$\ell^*$ also determines the width of the coherent backscattering cone in weak
localization~\cite{ALBADA-1985,MAYNARD-1986} and drives the phase-transition from diffusive to Anderson
localization~\cite{WIERSMA-2009}. An invariant quantity that does not depend on $\ell^*$ would thus be highly
surprising to the community involved in wave scattering through disordered media. Since, in addition, coherent effects
like weak or strong (Anderson) localization clearly fall outside the scope of a diffusive random walk model, one may
also expect that an invariance property simply does not exist when wave interference comes into play. As we will
demonstrate here explicitly, this expectation is clearly too pessimistic. Instead, we find that an invariant time and
length scale can also be defined for waves, even when they scatter non-diffusively as in the ballistic or in the
Anderson localization regime. The key insight that allows us to establish such a very general relation for the mean wave
scattering time is its connection to the density of states (DOS), which is the central quantity that stays invariant on
a level far beyond the scope of a diffusion approximation.

To describe wave transport in a disordered scattering medium without solving the full wave equation numerically
is a challenging task, which can be approached from many different
angles~\cite{Apresyan1996,Sheng1995,Akkermans2007}. As the first step, we will consider the radiative
transfer equation (RTE), which describes the
transport of an averaged radiation field through a disordered medium in the limit $k\,\ell_s\gg 1$, where
$k=\omega/c =2\pi/\lambda$ is
the wave number and $\ell_s$ is the scattering MFP~\cite{Chandrasekhar1960,CASE-1967}. In non-absorbing media as considered
here the scattering MFP $\ell_s$ is
connected to $\ell^*$ by the anisotropy parameter $g$ which measures the degree of forward-scattering
at a scattering event, $\ell_s=\ell^*\left(1-g\right)$. In its standard formulation where the RTE does not include wave
interference effects it should fully reproduce the predictions by Blanco~\emph{et al.}~from above. However, one can
enhance the scope of the RTE to include specific wave effects like the dispersion in a medium containing strongly
resonant scatterers such as atomic dipoles or Mie spheres~\cite{lagendijk1996review,PIERRAT-2009}.  In what follows, we
will consider identical, but randomly placed resonant and non-absorbing dipole scatterers described by a
polarizability $\alpha\left(\delta\right)=-4\pi/k^3\left[i+2\delta/\Gamma\right]^{-1}$, with $\delta=\omega-\omega_0$
the detuning with respect to the resonance frequency $\omega_0$, and $\Gamma$ the linewidth (modelling losses due
to scattering only)\footnote{The International System of Units is used throughout this article.}. This
specific expression is valid close to the resonance (\ie $\delta\ll\omega_0$) and ensures energy conservation (\ie the
optical theorem is fulfilled). In this case and for a dilute system such that ${\cal N}\lambda^3\ll 1$, where
${\cal N}$ is the density of scatterers, a dispersive form of the RTE can be derived from the Bethe-Salpeter equation
(an exact equation for the spatio-temporal auto-correlation function of the electric field)~\cite{PIERRAT-2009}:
\begin{multline}\label{eq:rte}
   \left[-\frac{i\Omega}{c}+\bm{u}\cdot\bm{\nabla}_{\bm{r}}+\mu_e\left(\delta,\Omega\right)\right]I\left(\bm{u},\bm{r},\delta,\Omega\right)=
\\
   =\frac{1}{4\pi}\mu_s\left(\delta,\Omega\right)\int I\left(\bm{u}',\bm{r},\delta,\Omega\right)\ud\bm{u}'
\end{multline}
where $\ud\bm{u}'$ stands for integration over the solid angle.  The specific intensity
$I\left(\bm{u},\bm{r},\delta,\tau\right)$ (also called spectral radiance) describes the radiative flux at position
$\bm{r}$, along direction $\bm{u}$, at frequency $\delta$ and at time $\tau$. It is defined from the Wigner transform of
the electric field. Equation~\eqref{eq:rte} is Fourier
transformed with respect to $\tau$ (with $\Omega$ being the conjugate Fourier variable). The expressions for the
extinction and scattering MFP are given by $\mu_e\left(\delta,\Omega\right)=-i {\cal N}
k/2\left[\alpha\left(\delta+\Omega/2\right)-\alpha^*\left(\delta-\Omega/2\right)\right]$ and
$\mu_s\left(\delta,\Omega\right)={\cal N}
k^4/(4\pi)\alpha\left(\delta+\Omega/2\right)\alpha^*\left(\delta-\Omega/2\right)$. The Boltzmann scattering MFP
$\ell_s\left(\delta\right)=\ell_0\left[1+4\delta^2/\Gamma^2\right]$, with $\ell_0 =\left[4\pi{\cal N}/k^2\right]^{-1}$ the
value at resonance ($\delta=0$), can be changed by varying the detuning $\delta$ or the linewidth $\Gamma$. Note
that the condition ${\cal N}\lambda^3\ll 1$ is then equivalent to $k\,\ell_s\gg 1$.

\begin{figure*}[!htbf]
   \centering
   \includegraphics[width=\linewidth]{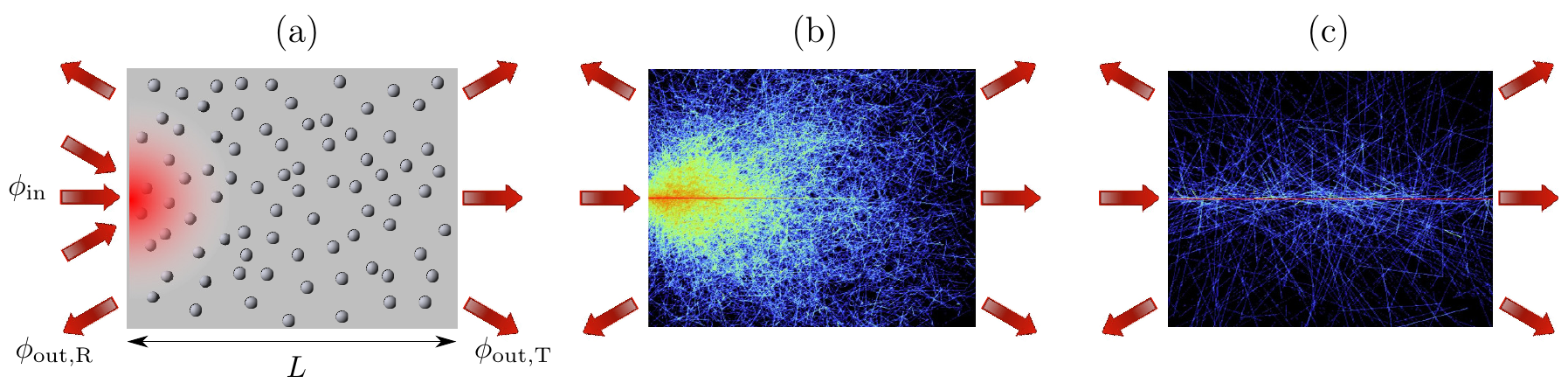}
   \caption{Sketch of the system and light trajectories. (a) Geometry of the 3D slab of length $L$
   investigated numerically using the RTE.  $\phiin$ is the incident flux and $\phioutr$, $\phioutt$ are 
   the reflected and transmitted fluxes, respectively ($\phiout=\phioutr+\phioutt$). (b),(c) Projections of light
   trajectories inside the system in the case of normal incidence illumination
   at a specific point on the system boundary (see incoming arrow) for two optical thicknesses (b) $b=10$
   and (c) $b=0.59$.}
   \label{fig:rte_slab1}
\end{figure*}

\begin{figure*}[!htbf]
   \centering
   \includegraphics[width=\linewidth]{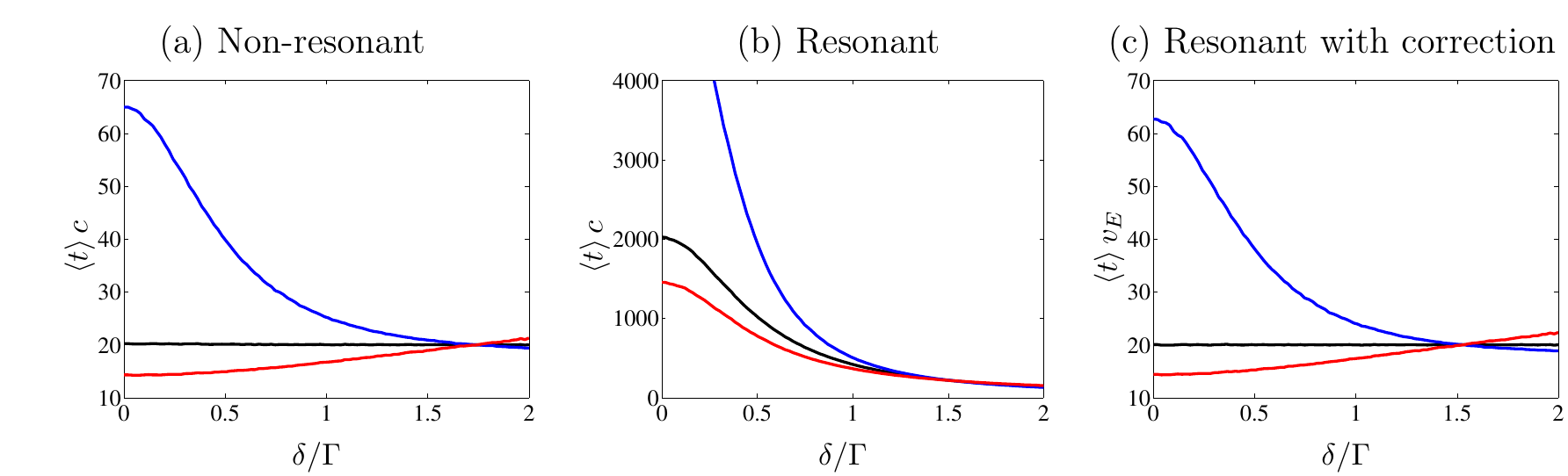}
   \caption{Ensemble-averaged length $\langle l \left(\delta\right)\rangle$ of light trajectories as obtained numerically using
   the RTE for (a) non-resonant and (b),(c) resonant scatterers in a 3D slab of width $L=10$ and optical thickness 
   at resonance $b_0=10$.
   Black/blue/red lines depict the values for all/transmitted/reflected trajectories, corresponding to the fluxes
   $\phiout$/$\phioutt$/$\phioutr$ in Fig.~\ref{fig:rte_slab1}. The values for $\langle l(\delta)\rangle$ were determined
   through the average time $\langle t\rangle$ multiplied by the speed of light $c$ in panels (a),(b) and by the energy
   velocity $v_E$ in panel (c). The renormalization with the energy velocity $v_E$ in the resonant case (c) yields the
   same universal value $\langle t\rangle v_E=2L$ as obtained by Blanco {\it et al}.~for the non-resonant case $\langle
   t\rangle c=2L$, see (a).  For $\delta=0$ and for $\delta=2\Gamma$, the optical thickness is $b=10$ and
   $b=0.59$, respectively, such that the above results range from the diffusive to the single-scattering regime.}
   \label{fig:rte_average_time}
\end{figure*}

On this basis we can evaluate the average time spent by light trajectories inside the medium by calculating the weighted temporal
average $\bra t\left(\delta\right)\ket=\int \tau\phiout\left(\delta,\tau\right)\ud
\tau/\int\phiout\left(\delta,\tau\right)\ud \tau$, where the weighting function $\phiout=\int_\Sigma \int_{2\pi}
I\left(\bm{u},\bm{r},\delta,\tau\right)\bm{u}\cdot\bm{n} \,\ud\bm{u}\,\ud^2\bm{r}$ is the outgoing flux at time $\tau$,
$\Sigma$ is the medium boundary and $\bm{n}$ the outward normal. In frequency domain, this expression can be cast in the
following compact form (see appendix~\ref{app:transport_average_time}):
\begin{equation}\label{eq:average_time_rte}
   \bra t\left(\delta\right)\ket=\frac{-i}{\phiout\left(\delta,\Omega=0\right)}
      \left.\frac{\partial\phiout\left(\delta,\Omega\right)}{\partial\Omega}\right|_{\Omega=0}\,,
\end{equation}
where we define $t=0$ as the time when the incident flux enters the medium. This expression is also convenient for a
numerical computation of $\bra t\left(\delta\right)\ket$ based on \eq{rte}. In the numerical simulation, we consider a
3D slab geometry of length $L$ with on-resonance optical thickness $b_0=L/\ell_0$, illuminated by an isotropic and uniform specific
intensity on its left interface only, see Fig.~\ref{fig:rte_slab1}(a). This corresponds to a situation where the
incident specific intensity does not depend on the point and direction of incidence (Lambert's cosine law is satisfied). Using a Monte-Carlo
scheme~\cite{PIERRAT-2009}, we solved Eq.~\eqref{eq:rte} without approximation and obtained the results plotted in
Fig.~\ref{fig:rte_average_time}. By tuning the linewidth $\Gamma$ of the scatterers, we can either simulate a
non-resonant medium in which the intensity spends most of the time between the scatterers ($\Gamma\ell_0\gg c$) or a
resonant medium where the transport time is dominated by intensity trapping inside the scatterers ($\Gamma\ell_0\ll c$). 

In the non-resonant case, see Fig.~\ref{fig:rte_average_time}(a), we recover the results by Blanco~\emph{et
al.}~\cite{BLANCO-2003} and find an average time $\bra t\left(\delta\right)\ket$ that is independent on the scattering
properties of the medium (\ie independent of the detuning $\delta$ that determines the scattering properties in the
present RTE calculation).  Moreover, we clearly see that the times associated to the reflected and transmitted parts of
the outgoing flux, that can be computed separately, strongly depend on $\delta$, showing that the invariance of the
average time $\bra t\left(\delta\right)\ket$ results from a delicate balance between reflection and transmission (in both
their intensities and time delays), as illustrated by light trajectories displayed in Fig.~\ref{fig:rte_slab1}(b),(c). Also
note that by varying the detuning $\delta$ from~$0$ to~$2$ in~Fig. \ref{fig:rte_average_time}(a), we effectively perform
a crossover from the diffusive to the single scattering regime. In the latter ($\delta\gtrsim 1.5$), the invariance of
$\bra t\left(\delta\right)\ket$ follows directly from the mean-chord-length theorem~\cite{CASE-1967}.

In case of a resonant medium, see Fig.~\ref{fig:rte_average_time}(b), the situation is substantially different.  The average
time $\bra t\left(\delta\right)\ket$ exhibits a significant dependence on $\delta$, and therefore on the scattering
properties of the medium.  As this result clearly falls outside the scope of the invariance relation derived by
Blanco~\emph{et al.}~\cite{BLANCO-2003}, the question arises, whether a new quantity can be defined that remains
invariant even in the limit of strongly dispersive scatterers.  To address this issue, we rewrite the average time $\bra
t\left(\delta\right)\ket$ in Eq.~\eqref{eq:average_time_rte} as the ratio of the total energy $U$ stored in the system
and the outgoing flux $\phiout$,
\begin{equation}\label{eq:average_time_rte2}
   \bra t\left(\delta\right)\ket=\frac{U\left(\delta,\Omega=0\right)}{\phiout\left(\delta,\Omega=0\right)}\,.
\end{equation}
As specified in more detail in the supplemental material, this relation measures $\bra t\left(\delta\right)\ket$ as the
time the stored energy $U$ takes to flow out of the medium with flux $\phiout=-\phiin$ (in stationary processes without
absorption or gain, the incoming and outgoing fluxes are balanced)~\cite{WINFUL-2003}.  Expressing $\phiout$ and $U$ in
terms of the specific intensity $\phiout\left(\delta,\Omega\right)=\int_\Sigma\int_{2\pi}
I\left(\bm{u},\bm{r},\delta,\Omega\right)\bm{u}\cdot\bm{n} \,\ud\bm{u}\,\ud^2\bm{r}$ and
$U\left(\delta,\Omega\right)=v_E^{-1}\left(\delta\right)\int_V\int_{4\pi}
I\left(\bm{u},\bm{r},\delta,\Omega\right)\ud\bm{u}\,\ud^3\bm{r}$, where $v_E(\delta)$ is the energy (or transport)
velocity~\cite{lagendijk1996review}, we obtain
\begin{multline}\label{eq:average_time_rte3}
   \bra t\left(\delta\right)\ket =\left[\frac{1}{v_E\left(\delta\right)}\int_V\int_{4\pi} I\left(\bm{u},\bm{r},\delta,\Omega=0\right)\ud\bm{u}\,\ud^3\bm{r}\right]
\\
      \times \left[
         \int_\Sigma\int_{2\pi} I\left(\bm{u},\bm{r},\delta,\Omega=0\right)\bm{u}\cdot\bm{n} \,\ud\bm{u}\,\ud^2\bm{r}
      \right]^{-1}.
\end{multline}
In this expression $V$ and $\Sigma$ are the volume and the external boundary of the medium, and $\bm{n}$ is the outward
normal. For a uniform and isotropic illumination on the surface (as assumed here), the specific intensity is uniform,
isotropic and independent on detuning inside the medium (a particular case of such a situation is blackbody
radiation)~\cite{Apresyan1996}.  As a result, Eq.~\eqref{eq:average_time_rte3} can be drastically simplified into $\bra
t\left(\delta\right)\ket =4V/\left[ \Sigma \,v_E\left(\delta\right)\right] $, which for a slab of thickness $L$ gives
$\bra t\left(\delta\right)\ket =2 L/v_E\left(\delta\right)$. This result turns out to be strikingly similar to the
invariance relation derived by Blanco~\emph{et al.}~\cite{BLANCO-2003}, the only difference being that in resonant media
the dispersive form of the energy velocity $v_E\left(\delta\right)$ comes into play. The expression of the energy
velocity for resonant scatterers can be determined explicitly~\cite{lagendijk1996review}, and takes the following form
(see appendix~\ref{app:transport_energy_velocity})
\begin{equation}\label{eq:energy_velocity}
   v_E\left(\delta\right)=\left[\frac{1}{c}+\frac{1}{\Gamma\ell_s\left(\delta\right)}\right]^{-1}\,.
\end{equation}
The energy velocity allows us to introduce an invariant length scale, $\bra l \ket=\bra t\left(\delta\right)\ket
v_E=4V/\Sigma$, which is independent on the scattering properties of the medium for both resonant and non-resonant
scattering (in the latter the energy velocity simply reduces to the constant velocity entering the random walk
formalism). To prove the correctness of this result, we plot the average length $\bra l \ket$ in
Fig.~\ref{fig:rte_average_time}(c) as obtained by renormalizing the numerical results for $\bra
t\left(\delta\right)\ket$ in Fig.~\ref{fig:rte_average_time}(b) with the analytical expression~\eqref{eq:energy_velocity}
of the transport velocity $v_E$.  We find that the resulting curve for $\bra l \left(\delta\right)\ket=\bra
t\left(\delta\right)\ket v_E$ is, indeed, independent on the detuning $\delta$, with a constant value $\bra l \ket=2L$.
This result is all the more remarkable as the average lengths associated with either the transmitted or the reflected part
of the flux display a strong dependence on the scattering properties in the same regime. This again shows that the
invariance of the average length $\bra l \ket$ results from a subtle balance between reflection and transmision. 

Whereas the above extension of the RTE allowed us to find a new invariant quantity for the case of scattering in a
disordered medium with resonant scatterers, the ansatz of the RTE itself is intrinsically restricted to the limit
$k\ell_s\gg 1$.  The opposite limit, where the wave length $\lambda$ is comparable to or even larger than the mean free path
$\ell_s$, is thus not covered by our foregoing considerations. As in this strongly scattering limit wave interference can
lead to a complete halt of wave diffusion in terms of Anderson localization, the question arises whether localization
will lead to a deviation from the above invariance property or not. One could expect such a deviation, e.g., on the
grounds that localization prevents scattering states to explore the entire scattering volume $V$ of the system.
Correspondingly, the volume $V$ and the surface $\Sigma$ appearing in the invariance relation $\bra
t\left(\delta\right)\ket =4V/\left[ \Sigma \,v_E\left(\delta\right)\right] $ might then have to be rescaled with the
localization length $\xi$. 

To explore this question in detail we will now work with the full wave equation in two dimensions which, for stationary
light scattering, is given in terms of the Helmholtz equation
\begin{equation}\label{eq:Helmholtz}
   \left[ \Delta + n\!\left(x,y\right)^2 k^2 \right] \psi\!\left(x,y\right) = 0\,.
\end{equation} 
The linear dispersion $k=\omega/c$ will allow us to use $k$ and $\omega$ interchangeably. 
In the situations we study here, the disorder scattering is induced by the spatial variations of the
static refractive index $n\!\left(x,y\right)$. To evaluate the dwell time of a stationary scattering eigenstate of this
equation (with well-defined wave number $k$) inside a given spatial region one can conveniently use the so-called
Wigner-Smith time-delay operator\footnote{One can show that the quantity measured by the Wigner-Smith time-delay
operator is equal to the dwell time (Eq.~\eqref{eq:average_time_rte2}) if the frequency dependence of the coupling
between the scattering region and its surrounding becomes negligible~\cite{Sokolov1997}. This is the case in the systems
considered here.} 
\begin{equation}\label{eq:WignerSmith}
  Q(\omega)=-i\, S^{-1}\frac{d S}{d \omega}\,,
\end{equation} 
originally introduced by Wigner in nuclear scattering theory~\cite{Wigner1955} (and extended by Smith to multi-channel
scattering problems~\cite{Smith1960}). Here the $\omega$-dependent scattering matrix $S$, evaluated at the external
boundary  $C$ of the considered region, contains all the complex transmission and reflection amplitudes that connect in-
and outgoing waves in a suitable mode basis.  To obtain also here the average time associated with wave scattering we
take the trace of $Q$ and divide by the number $N(\omega)$ of incoming scattering channels, $\langle
t(\omega)\rangle={\rm Tr} \! \left[ Q(\omega) \right] / N(\omega)$.

\begin{figure*}[!htbf]
   \centering
   \includegraphics[width=\linewidth]{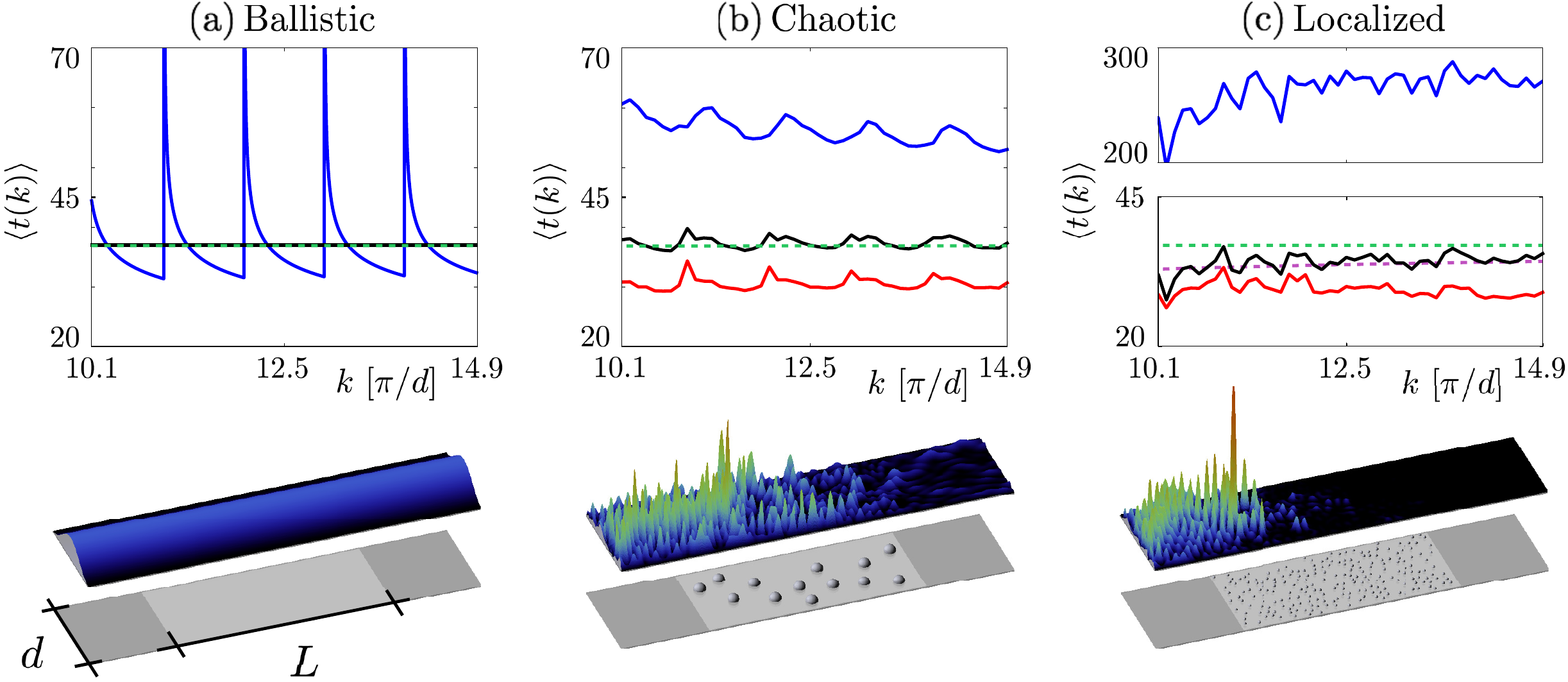}
   \caption{Total average dwell time $\left\langle t \! \left( k \right) \right\rangle$ (black line), transmission delay
   time (blue line), and reflection delay time (red line) for (a) ballistic scattering through a clean waveguide as well as for (b) chaotic scattering
   through a disordered waveguide with 13 circular obstacles of radius $r=0.06 \, d$ and (c) Anderson localized transport through 
   a disordered waveguide with  211 obstacles of $r=0.015 \, d$ (see appendix~\ref{app:average_time}
   for a definition of transmission and reflection delay times). The geometrical parameters were chosen such that
   all three waveguides have the same width $d$ and the same effective scattering area $A=2.35\,d^2$. 
   The wavenumber was scanned between $k=10.1 \, \pi/d$
   and $k=14.9 \, \pi/d$ in all three cases.  For the clean waveguide in (a) the transmission is perfect, thus the reflection times
   are strictly zero. The average for the total dwell time (black line) is taken here over the entire wavenumber
   interval shown and coincides with the estimate of Blanco {\it et
   al.}~$\langle t(k)\rangle=\pi A/(Cc)$ (green dashed line). 
   For the disordered systems in (b),(c), the averages were taken over (b) 250 and (c) 2500 different random configurations,
   respectively. Whereas for the chaotic scattering case (b) the results for the average dwell time agree well with the
   random walk prediction (dashed green line), a systematic deviation is observed for the case of strong disorder (c). 
   Here, very good agreement is found with the estimate for the average dwell 
   time according to the corrected Weyl
   estimate, Eq.~\eqref{eq:Weyl-Boundary} (purple dashed line).  In the lower panels, the intensity of wave functions
   injected in the lowest-order mode is shown for a specific configuration of scatterers (see grey spheres) embedded in
   the scattering area (light grey domain in the middle). The flux is incoming from the left and can be transmitted (to
   the right) or reflected (to the left) through the perfect waveguides attached on both sides (see dark grey areas).}
   \label{fig:AveragePathWave}
\end{figure*}

To evaluate the average time $\langle t(\omega)\rangle$ from above, we performed numerical
simulations on a
two-dimensional scattering region of rectangular shape, attached to perfect semi-infinite waveguides on the left and
right (see illustrations in the lower panels of Fig.~\ref{fig:AveragePathWave}). Accordingly, the correct number of
scattering channels $N(\omega)$ is given by the total number of flux-carrying modes in both waveguides.  Impenetrable
and non-overlapping circular scatterers are randomly placed inside the scattering region and in between them the
refractive index is kept constant, $n(x,y)=1$.  The scattering matrix and the corresponding scattering states for this
system are calculated by solving the Helmholtz Eq.~\eqref{eq:Helmholtz} on a finite-difference grid, using the advanced
modular recursive Green's function method~\cite{Rotter2000,Libisch2012}.  In Fig.~\ref{fig:AveragePathWave} we display
our numerical results for different degrees of disorder: In subfigure~(a), we show the results obtained for an empty
scattering region, corresponding to the ballistic transport regime. In subfigure~(b), the case with altogether $13$
scatterers is shown, for which already a strong reduction of transmission is observed. The distribution of the
transmission eigenvalues $P(\tau)$ follows here very well the predictions of Random Matrix Theory  for the regime of
chaotic scattering (see appendix~\ref{app:eigenvalues}). Finally, in subfigure~(c), we increased the degree of disorder even more
(placing altogether $211$ scatterers) such as to enter the regime of Anderson localization. Here the distribution of
transmission eigenvalues agrees very well with the predictions for the case when Anderson localization suppresses all
but a single transmission eigenchannel (see appendix~\ref{app:eigenvalues})~\cite{Gopar2010, Pena2014}.  To make all three cases
easily comparable with each other, the different geometries all have the same scattering area $A$, which, for
ballistic scattering is the entire rectangular region between the leads, whereas for the other two cases the area
occuppied by the impenetrable scatterers is not part of $A$.

Based on the above identification of the different transport regimes that our model system can be in, we investigate now
the corresponding results for the average time $\bra t(\omega)\ket$ which we get for each of these limits (see panels in
Fig.~\ref{fig:AveragePathWave}). In the ballistic limit [see panel (a)], we see that the average time, plotted as a
function of the incoming wavenumber $k$, shows pronounced periodic enhancements around the random walk
prediction by Blanco {\it et al.}~$\bra t \ket = \pi A/(C v)$. The peaks of these fluctuations can be identified with
those positions in $k=k_n = n \, \pi/d$, where a new transverse mode opens up in the waveguide of width $d$. To
understand why these mode openings cause an increase in the scattering dwell time, we resort to a fundamental
connection between the average dwell time $\langle t\rangle$ and the density of states (DOS) $\rho(k)$.
This relation, $\rho(k)=N(k) c \bra t(k) \ket /(2\pi) = c \, \rm{Tr} \! \left( Q\right) / (2\pi)$, was first put forward
by Birman, Krein, Lyuboshitz and Schwinger in the context of quantum electrodynamics and nuclear scattering theory and
has meanwhile been used in a variety of different contexts~\cite{Schwinger1951, Krein1962, Krein1962a, LYUBOSHITZ-1977,
Birman1992, Time1995, Fyodorov1997, Souma2002, Yamilov2003, Genack2003, Davy2014}.  Since, in the ballistic regime, each
individual incoming mode corresponds to a one-dimensional scattering channel with, correspondingly, an associated square
root singularity in the DOS, $\rho_n(k)=[L/\!\left(2\pi \right)] \, k/\!\sqrt{k^2-k^2_n}$ for 
$k>k_n$, we can successfully
explain the observed oscillations as coming from the successive openings of new waveguide modes. Evaluating the total 
DOS based on a sum of individual mode contributions, $\rho(k)=\sum_n^N \rho_n(k)$, and using the above connection to 
the average time yields identical results to those shown in 
Fig.~\ref{fig:AveragePathWave}(a). This demonstration also allows us to show that the time, averaged over an
interval of $k$ that is larger than the distance between successive mode openings, converges exactly to the prediction by
Blanco \emph{et al.} Quite remarkably, we find in this sense that the estimate from the mean-chord-length theorem and,
correspondingly, the random walk prediction also holds, on average, for ballistic wave scattering in a system without
any disorder.

Moving next to the disordered system in  Fig.~\ref{fig:AveragePathWave}(b), we see that the presence of the disorder
strongly reduces the above mode-induced fluctuations, leaving the frequency-average value of time unchanged. To explain
this result, the DOS clearly needs to be estimated differently here than in the ballistic case of uncoupled waveguide
modes.  Also, since the disorder leads to system and frequency specific fluctuations of the DOS, we are looking here for
an estimate for the ensemble and frequency averaged DOS. To obtain this quantity, we invoke a result first put forward
by Weyl in 1911 \cite{Weyl1911}, who estimated that the average DOS in the asymptotic limit of $\omega\to\infty$
satisfies the following universal law, $\rho(\omega)=A\omega/(2\pi c^2)$, now called the Weyl
law~\cite{arendt_weyls_2009}. Putting this estimate into the formula relating the average time with the average DOS, we
obtain $\bra t(\omega) \ket =2\pi\rho(\omega)/N(\omega)=A\omega/[c^2 N(\omega)]$. The $\omega$-dependent number of
incoming channels is given as an integer-valued step-function $N(\omega)=\floor*{2\omega d/(c \pi)}$.  When smoothing
over the steps in this function, i.e, $N(\omega)\approx 2\omega d/\left(c\pi\right)-0.5$, we arrive at the result $\bra
t(\omega) \ket =2\pi\rho(\omega)/N(\omega) \approx \pi A / (2 d c) = \pi A / (C c)$. This relation, which is very
accurately satisfied  by our numerical results, thus confirms the validity of the diffusive  random walk prediction by
Blanco {\it et al.}~also for disordered wave scattering. Because the above relation for the average dwell time is notably
independent of $\ell^*$, transmission and reflection times for waves, which do strongly depend on $\ell^*$, need to fully
counter-balance each other. 

Does this invariance of the average scattering time also persist in the strongly scattering limit, when Anderson
localization sets in? Our numerical results shown for this case in Fig.~\ref{fig:AveragePathWave}(c) display a small but
apparently systematic frequency dependence of the average time $\bra t(\omega)\ket$ which increasingly deviates from the
result by Blanco~\emph{et al.}~for decreasing frequencies $\omega$.  Since the numerical calculations are very
challenging and the frequency derivative appearing in Eq.~\eqref{eq:WignerSmith} can reach very large values for highly
localized scattering states, we first tested the accuracy of our simulations by evaluating $\bra t(\omega)\ket$ also
through explicit dwell time calculations. In analogy to Eq.~\eqref{eq:average_time_rte2}, the expression for the dwell
time in case of the Helmholtz Eq.~\eqref{eq:Helmholtz} is given by $ t_m = \int_A \psi_m^\star n^2 \psi_m \, d^2\bm{r}
\, / \phi_{m,\textrm{in}}$, where $\psi_m$ is the wave function of the $m$-th scattering channel and
$\phi_{m,\textrm{in}}$ is the corresponding total (stationary) incoming flux. The average dwell time is then given by
$\left\langle t\left(\omega\right) \right\rangle = \Sigma^N_m t_m\left(\omega)\right) / N \left(\omega\right)$. The
results obtained in this way are practically indistinguishable from Fig.~\ref{fig:AveragePathWave}(c).
To explain this robust deviation from the result by Blanco~\emph{et al.}~\cite{BLANCO-2003}, we thus
have a more careful look on the Weyl estimate which, in addition to the leading order term which we used above, also
features a next-order correction proposed by Weyl~\cite{Weyl1913,arendt_weyls_2009},  $\rho \left(\omega\right) =
\left[A\omega/c^2 + \left(C-B\right)/\!\left(2 c\right) \right] / \!\left(2 \pi\right)$. This correction
involves not only the scattering area $A$, but also the internal boundary of the scattering region $B$ which is notably
different from the external boundary $C$ through which waves can scatter in and out. The internal boundary $B$ in case
of our waveguide system under study is given by $B = 2 L + B_o$, where $B_o$ is given by the total circumference of the
scatterers.  The open boundary conditions along the external boundary $C$ were approximated with Neumann boundary
conditions, which contribute with the opposite sign as the Dirichlet boundary conditions on the surface of the waveguide
and of the scatterers. In systems with a small boundary-to-area ratio this next-order correction of the Weyl law is negligible. Since,
however, the number of scatterers which we have placed inside the system (from $0$ in the ballistic case, to $13$ in the
chaotic case, to $211$ in the localized case) increases this ratio, the additional boundary term in the Weyl law may become
important here. To check this explicity, we re-evaluate the expression for the average dwell time $\bra t(\omega)\ket$
from above when adding this correction, leading us to
\begin{equation}\label{eq:Weyl-Boundary}
  \left\langle t \left(\omega\right) \right\rangle = \frac{1}{c^2 N \! \left(\omega\right)} \left[ A \, \omega + \frac{\left(C-B\right)}{2} c \right].
\end{equation}
A comparison of this analytical formula with the numerical results, see Fig.~\ref{fig:AveragePathWave}(c), 
yields excellent agreement and indicates that the
observed deviation from the prediction by Blanco {\it et al.}~stems from the comparatively large boundary of the many
small scatterers which we placed inside the scattering region. We emphasize at this point that this correction to the
Blanco estimate only contains the boundary values $B$ and $C$ as additional input and remains entirely independent of
any quantities that characterize the scattering process itself, like $\ell^*$ or the localization
length $\xi$. This insight is of considerable importance, since it means that Eq.~\eqref{eq:Weyl-Boundary}  defines a
new invariant quantity that is independent of the scattering regime we are in and thus  accurately matches our numerical
results for the average time in the ballistic, chaotic and localized limit. This invariant quantity for waves deviates
from the prediction by Blanco~\emph{et al.}~\cite{BLANCO-2003} only through an additional term originating in the fact
that waves feel the boundary of a scattering region already when being close to it on a scale comparable with the wave
length. We speculate that additional wave corrections to the result by Blanco and Fournier may arise when waves have
access to a larger scattering area $A$ than classical particles through the process of tunneling.

In summary, we have derived a universal invariance property for wave transport through disordered media.  The invariance
of the averaged path length or averaged time spent by a wave in an open finite medium has been established based on
scattering theory.  In the appropriate limit of diffusive and non-resonant media, the random-walk picture is recovered,
and the result coincides with the expression of the averaged path-length initially established by Blanco~\emph{et
al.}~\cite{BLANCO-2003}. Our work confers to this invariance property a degree of universality that extends its
implications far beyond applications of random-walk theory.  This extension to waves opens up new possible applications
in optics, acoustics, seismology, or radiofrequency technologies, where  propagation in complex media is the subject of
intense research~\cite{ishimaru1978wave}.  Indeed, in the context of wave transport through disordered media, most spatial or
temporal observables scale with $\ell^*$, and the invariance property derived in this work is
particularly counterintuitive and rich in implications.  It should find applications in imaging, communications, or
light delivery, for instance to generate enhanced light-matter interaction within a certain volume by controlled light
deposition, or to design specific structures to enhance light harvesting for solar cells~\cite{vynck2012,
yablonovitch1982intensity}. Consider here for example that the above invariance property allows us to estimate the time that
waves need to transit through a given medium based on a measurement of only the reflected portion of incoming waves and
an \emph{a priori} knowledge of the sample geometry. Particularly intriguing in our eyes is the possibility to get access,
through Eq.~\eqref{eq:Weyl-Boundary}, to the internal surface $B$ of scatterers embedded in a scattering medium through
a time-resolved transport experiment. Such an approach could go as far as to measure the fractal dimension of the scatterer surface by
linking our results with the Berry-Weyl conjecture~\cite{Berry1979,Lapidus1991}.

An extension of our findings to media with gain and loss~\cite{cao2003lasing, chong2011hidden} should also be of interest,
both from a theoretical and an applied standpoint. Our study should also be very relevant to the field of wave control,
which has recently emerged as a powerful paradigm for light manipulation and delivery in complex
media~\cite{mosk2012controlling}, showing for instance that suitably shaped wavefronts can deliver light at a specific time and
position~\cite{aulbach2011control,vellekoop2008universal,Rotter2011}.  Finally, let us point out that although we only
studied here 3D slab and 2D waveguide geometries with uncorrelated disorder, the invariance property, thanks to its
connection to the DOS, is very general and should apply to a wide range of geometries and excitation strategies, as well
as to non-uniform scattering properties, biological tissues, and correlated disorder, from partially ordered to entirely ordered
system such as Levy glasses or photonic crystals~\cite{ barthelemy2008levy, wiersma2013disordered}. An experimental
demonstration of the discussed invariance property should be within reach, in particular in optics where
time-resolved techniques and sensitive detectors are available.

% Acknowledgments
\begin{acknowledgments}
   This work was supported by LABEX WIFI (Laboratory of Excellence ANR-10-LABX-24) within the French Program
   ``Investments for the Future'' under reference ANR-10-IDEX-0001-02 PSL$^{\ast}$.  PA, AH and SR are supported by the
   Austrian Science Fund (FWF) through Projects NextLite F49-10, and I 1142-N27 (GePartWave).  SG is funded
   by the European Research Council (grant no. 278025), and  would like to thank St\'ephane Hallegatte for pointing out
   the result of Blanco {\it et al.}~and for an initial exchange of ideas. We also thank Jacopo Bertolotti, Florian
   Libisch, Romolo Savo and Jolanda Schwarz for fruitful discussions as well as the administration of the Vienna
   Scientific Cluster (VSC) for granting us access to computational resources.
\end{acknowledgments}

% Appendix
\appendix

\section{Transport equation and energy velocity}\label{app:transport_energy_velocity}
% ==============================================

We recall that the transport equation for a resonant scattering system is given by~\cite{PIERRAT-2009}
\begin{multline}\label{RTE}
   \left[-\frac{i\Omega}{c}+\bm{u}\cdot\bm{\nabla}_{\bm{r}}+\mu_e\left(\omega,\Omega\right)\right]
   I\left(\bm{u},\bm{r},\omega,\Omega\right)
\\
   =\frac{1}{4\pi}\mu_s\left(\omega,\Omega\right)\int
   I\left(\bm{u}',\bm{r},\omega,\Omega\right)\ud\bm{u}'
\end{multline}
where $I$ is the specific intensity, proportional to the radiative flux at position $\bm{r}$, in direction $\bm{u}$,
at frequency $\omega$ and at time $\tau$ ($\Omega$ in frequency domain). $c$ is the speed of light in vacuum.
$\mu_e\left(\omega,\Omega\right)$ and $\mu_s\left(\omega,\Omega\right)$ are coefficients given by
\begin{align}
   \mu_e\left(\omega,\Omega\right) & = \frac{-i{\cal N} k}{2}
      \left\{\alpha\left(\omega+\frac{\Omega}{2}\right)-\alpha^*\left(\omega-\frac{\Omega}{2}\right)\right\}
\\
   \textrm{and }\mu_s\left(\omega,\Omega\right) & = \frac{{\cal N} k^4}{4\pi} \alpha\left(\omega+\frac{\Omega}{2}\right)\alpha^*\left(\omega-\frac{\Omega}{2}\right)
\end{align}
where $\alpha$ is the polarizability of a point scatterer (dipole) and ${\cal N}$ the density.
To deal with resonant scatterers, we have chosen to write $\alpha$ in the form
\begin{equation}
   \alpha\left(\omega\right)=\frac{-4\pi}{k^3}\frac{1}{i+2\left(\omega-\omega_0\right)/\Gamma}
\end{equation}
where $k=\omega/c$. This expression fulfills the optical theorem (energy conservation), no losses by
absorption are present. Defining the detuning by $\delta=\omega-\omega_0$, the scattering length is thus given by
\begin{equation}\label{ell}
   \boxed{
      \ell\left(\delta\right)=\ell_0\left[1+\frac{4\delta^2}{\Gamma^2}\right]
   }
\end{equation}
where $\ell_0=\left[4\pi{\cal N}/k_0^2\right]^{-1}$ is the scattering length at the resonant frequency $\omega_0$.

Integrating Eq.~\eqref{RTE} over the directions (first moment), it is possible to derive a conservation equation
linking the energy density $u$ and the radiative flux vector $\bm{\phi}$ defined as follows:
\begin{align}
   u\left(\bm{r},\omega,\Omega\right) & =\frac{1}{v_E}\int I\left(\bm{u},\bm{r},\omega,\Omega\right)\ud{\bm{u}},
\\
   \bm{\phi}\left(\bm{r},\omega,\Omega\right) & =\int I\left(\bm{u},\bm{r},\omega,\Omega\right)\bm{u}\ud{\bm{u}}.
\end{align}
We obtain:
\begin{multline}
   \left[-\frac{i\Omega}{c}+\left\{\mu_e\left(\omega,\Omega\right)-\mu_s\left(\omega,\Omega\right)\right\}\right]
      v_E u\left(\bm{r},\omega,\Omega\right)
\\
   +\bm{\nabla}_{\bm{r}}\cdot\bm{\phi}\left(\bm{r},\omega,\Omega\right)=0.
\end{multline}
To identify with a conservation equation of the form
\begin{equation}\label{balance}
   -i\Omega u\left(\bm{r},\omega,\Omega\right)+\bm{\nabla}_{\bm{r}}\cdot\bm{\phi}\left(\bm{r},\omega,\Omega\right)=0,
\end{equation}
the energy velocity should read
\begin{equation}
   \frac{1}{v_E\left(\omega,\Omega\right)}=\frac{1}{c}
      +\frac{i}{\Omega}\left\{\mu_e\left(\omega,\Omega\right)-\mu_s\left(\omega,\Omega\right)\right\}.
\end{equation}
Taking the limit $\Omega\to 0$, we finally obtain
\begin{equation}\label{v_E}
   \boxed{
      \frac{1}{v_E\left(\delta\right)}=\frac{1}{c}+\frac{1}{\Gamma\ell\left(\delta\right)}
   }.
\end{equation}

\section{Transport equation and average time}\label{app:transport_average_time}
% ===========================================

The average time is defined by
\begin{equation}
   \bra t\left(\delta\right)\ket=\bra \tout\left(\delta\right)\ket-\bra \tin\left(\delta\right)\ket
\end{equation}
where the incoming and outgoing average times are given by
\begin{align}
   \bra \tin\left(\delta\right)\ket & =\frac{\int \tau\phiin\left(\delta,\tau\right)\ud \tau}{\int \phiin\left(\delta,\tau\right)\ud \tau}
\\
   \bra \tout\left(\delta\right)\ket & =\frac{\int \tau\phiout\left(\delta,\tau\right)\ud \tau}{\int \phiout\left(\delta,\tau\right)\ud \tau}
\end{align}
and $\phi_{\mathrm{in},\mathrm{out}}\left(\delta,\tau\right)$ are the input/output fluxes at time $\tau$ and for
a detuning $\delta$. In frequency domain, this reads
\begin{equation}\label{average}
   \boxed{
      \bra t_{\mathrm{in},\mathrm{out}}\left(\delta\right)\ket = \frac{-i}{\phi_{\mathrm{in},\mathrm{out}}\left(\delta,\Omega=0\right)}
         \left.\frac{\partial\phi_{\mathrm{in},\mathrm{out}}\left(\delta,\Omega\right)}{\partial\Omega}\right|_{\Omega=0}
   }.
\end{equation}
By integrating Eq.~\eqref{balance} over the volume of the system we get
\begin{equation}
   i\Omega \int_V u\left(\bm{r},\delta,\Omega\right)\ud^3\bm{r}
      =\int_V \bm{\nabla}_{\bm{r}}\cdot\bm{\phi}\left(\bm{r},\delta,\Omega\right)\ud^3\bm{r}
\end{equation}
and using the divergence theorem we find
\begin{multline}
   i\Omega \int_V u\left(\bm{r},\delta,\Omega\right)\ud^3\bm{r}
      =\int_{\Sigma} \bm{\phi}\left(\bm{r},\delta,\Omega\right)\cdot\bm{n}\,\ud^2\bm{r}=\phi\left(\delta,\Omega\right)
\\
      =\phiin\left(\delta,\Omega\right)+\phiout\left(\delta,\Omega\right).
\end{multline}
As the system is not absorbing, the stationary outgoing flux is given by
$\phiout\left(\delta,\Omega=0\right)=-\phiin\left(\delta,\Omega=0\right)$ and the Taylor expansion of the fluxes writes
\begin{equation}
   \phi_{\mathrm{in},\mathrm{out}}\left(\delta,\Omega\right)\sim\phi_{\mathrm{in},\mathrm{out}}\left(\delta\right)
      +\Omega\left.\frac{\partial \phi_{\mathrm{in},\mathrm{out}}\left(\delta,\Omega\right)}{\partial\Omega}\right|_{\Omega=0}.
\end{equation}
Thus, the total stationary energy inside the system writes
\begin{multline}
   \int_V u\left(\bm{r},\delta,\Omega=0\right)\ud^3\bm{r}=\\
      -i\left.\frac{\partial \phiin\left(\delta,\Omega\right)}{\partial\Omega}\right|_{\Omega=0}
      -i\left.\frac{\partial \phiout\left(\delta,\Omega\right)}{\partial\Omega}\right|_{\Omega=0}
\end{multline}
and the average time becomes
\begin{equation}\label{average2}
   \boxed{
      \bra t\left(\delta\right)\ket = \frac{U\left(\delta,\Omega=0\right)}{\phiout\left(\delta,\Omega=0\right)}
   }
\end{equation}
where $U$ is the total energy stored within the system.  Using the definition of the energy density, we find that the
average time renormalized by the energy velocity is given by
\begin{multline}
   \bra t\left(\delta\right)\ket v_E = 
      \left[\int_{\Sigma}\int_{2\pi} I\left(\bm{u},\bm{r},\omega,\Omega=0\right)\bm{u}\cdot\bm{n}\ud\bm{u}\ud^2\bm{r}\right]^{-1}
\\
      \times\int_V\int_{4\pi} I\left(\bm{u},\bm{r},\omega,\Omega=0\right)\ud\bm{u}\ud^3\bm{r}.
\end{multline}
This quantity can be seen as the average length of the random walk process inside the system and as it depends only on
the specific intensity for a given frequency, this is the right quantity that should be conserved whatever the detuning.
Indeed, if we illuminate the system with an isotropic specific intensity $I_0$ at each point of the boundary, the only
solution is $I=I_0$ inside the system and the average length reads
\begin{equation}
   \bra t\left(\delta\right)\ket v_E = \frac{4\pi V I_0}{\pi \Sigma I_0} = \frac{4V}{\Sigma}
\end{equation}
where $\Sigma$ and $V$ are the surface and the volume of the system, respectively.

\section{Average Transmission and Reflection Delay Times}\label{app:average_time}
% =======================================================

The average total delay time in scattering systems described by a
wave equation such as the Helmholtz equation can conveniently be written
as the trace of the time delay operator $Q$ divided by the total
number of open scattering channels $N$ (see also main text). Using the scattering amplitudes stored in the scattering
matrix $S$, we can rewrite the corresponding
expression as follows
\begin{equation}
\left\langle t\right\rangle =\frac{1}{N}\,{\rm
Tr}\!\left(Q\right)=\frac{1}{N}\left(\sum_{m,n}^{N}\left|S_{mn}\right|^{2}\frac{\ud\varphi_{mn}}{\ud\omega}\right),\label{eq:av_total_time}
\end{equation}
where $S_{mn}=\left|S_{mn}\right|e^{i\varphi_{mn}}$ is the complex
scattering amplitude connecting the $n$-th incoming and the $m$-th
outgoing channel. For the 2-port systems we study, the scattering
matrix can formally be decomposed into four distinct blocks,
\begin{equation}
   S=\left(\begin{array}{cc}
      r & t^{\prime}\\
      t & r^{\prime}
   \end{array}\right),
\end{equation}
the matrices $r$ and $t$ contain the elements associated with reflection and transmission for injection through the left
waveguide, respectively.  The primed quantities contain the corresponding elements for injection from the right. Using
this division into reflected and transmitted parts, we can define the average total transmission $\left\langle T_{\rm tot}\right\rangle
$ and reflection $\left\langle R_{\rm tot}\right\rangle $ according to
\begin{multline}
   \left\langle T_{\rm tot}\right\rangle
   =\frac{1}{N}\left(\sum_{m,n}^{N/2}\left|t_{mn}\right|^{2}+\left|t_{mn}^{\prime}\right|^{2}\right)
\\
   =1-\frac{1}{N}\left(\sum_{m,n}^{N/2}\left|r_{mn}\right|^{2}+\left|r_{mn}^{\prime}\right|^{2}\right)=1-\left\langle
   R_{\rm tot}\right\rangle .
\end{multline}
The effective number of transmitting channels then evaluates to $N_{T}=\left\langle T_{\rm tot}\right\rangle N$ and analogously
the effective number of reflected channels is $N_{R}=\left\langle R_{\rm tot}\right\rangle N$.  Very similar to
Eq.~\eqref{eq:av_total_time}, we can then finally define the average transmission time $\left\langle t_{T}\right\rangle
$ and the average reflection time $\left\langle t_{R}\right\rangle $ as
\begin{equation}
   \left\langle t_{T}\right\rangle
   =\frac{1}{N_{T}}\left(\sum_{m,n}^{N/2}\left|t_{mn}\right|^{2}\frac{\ud\varphi_{mn}^{t}}{\ud\omega}+\left|t_{mn}^{\prime}\right|^{2}
      \frac{\ud\varphi_{mn}^{t^{\prime}}}{\ud\omega}\right),\label{eq:av_trans_time}
\end{equation}
and
\begin{equation}
   \left\langle t_{R}\right\rangle =\frac{1}{N_{R}}\left(\sum_{m,n}^{N/2}\left|r_{mn}\right|^{2}
      \frac{\ud\varphi_{mn}^{r}}{\ud\omega}+\left|r_{mn}^{\prime}\right|^{2}\frac{\ud\varphi_{mn}^{r^{\prime}}}{\ud\omega}\right),\label{eq:av_refl_time}
\end{equation}
with, e.g., $r_{mn}=\left|r_{mn}\right|^{2}e^{i\varphi_{mn}^{r}}$ denoting a complex reflection amplitude from left to
left. Note that the properly weighted sum of the times~\eqref{eq:av_trans_time} and~\eqref{eq:av_refl_time} add up to
the average total time,
\begin{equation}
   \left\langle t\right\rangle =\left\langle T_{\rm tot}\right\rangle \left\langle t_{T}\right\rangle +\left\langle R_{\rm tot}\right\rangle \left\langle t_{R}\right\rangle .
\end{equation}

\section{Statistical signature for the Chaotic and the Localized Regime}\label{app:eigenvalues}
% ======================================================================

\begin{figure*}[!htbf]
   \centering
   \includegraphics[width=0.7\linewidth]{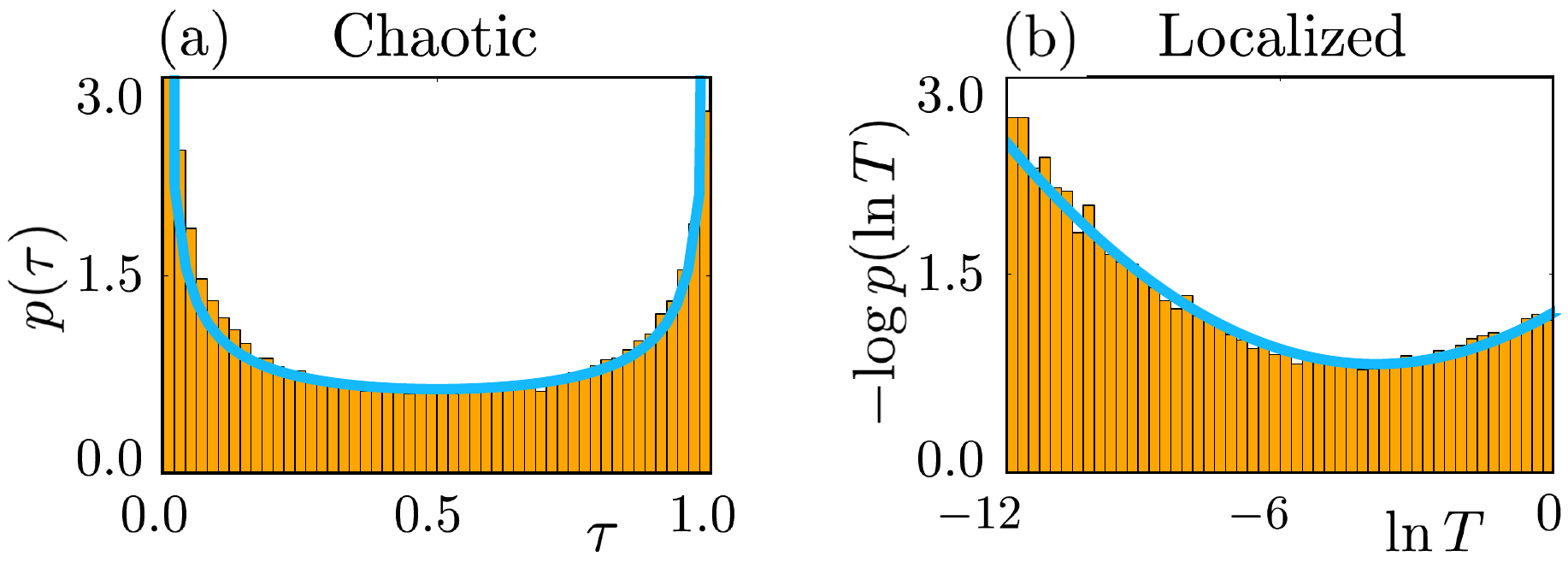}
   \caption{Transmission statistics for different transport regimes. (a) Distribution of the $t^{\dagger}t$ eigenvalues
   $\tau$ for chaotic scattering (orange bars), compared with the prediction, Eq.~\eqref{eq:chaos-dist} (light blue
   line).  (b) Distribution of the transmission $T$ for localized scattering compared with the prediction based on
   Eq.~\eqref{eq:loc-dist}. To produce the histograms, $k$ was scanned between $k=12.1 \pi/d$ and $k=12.9\pi/d$ and
   $1000$ scatterer configurations were considered for each of the cases (a),(b) [in (a) only values $0.01<\tau<0.99$
   were considered for the histogram since for very small and very large values of $\tau$ deviations from
   Eq.~\eqref{eq:chaos-dist} are expected~\cite{Rotter2007}].}
   \label{fig:Distributions}
\end{figure*}

In the main text, we discuss systems featuring ballistic, chaotic and localized wave scattering, respectively. The
corresponding scattering regime is determined by the number and size of impenetrable obstacles we placed inside the
scattering region and can be characterized through the regime-specific transmission statistics.  For the ballistic
system, transmission is perfect in our case, since without any scatterers we are dealing with a perfectly transmitting
waveguide. In order to verify that the scattering in the systems containing a finite number of obstacles is chaotic and
localized, respectively, we check whether the transmission statistics follow the respective predictions. For that
purpose, we calculated the eigenvalues $\tau_{i}$ of the matrix $t^{\dagger}t$, where $t$ is the transmission matrix.
For chaotic dynamics, the $\tau_{i}$ follow the bimodal distribution \cite{Baranger1994,Jalabert1994,Beenakker1997}
\begin{equation}
   p\!\left(\tau\right)=\frac{1}{\pi\sqrt{\tau\left(1-\tau\right)}}.\label{eq:chaos-dist}
\end{equation}
In a sample with Anderson localization only one single transport channel dominates the transmission~\cite{Pena2014},
such that the transmission, $T=\Sigma_{i=1}^{N/2}\tau_{i}\approx\tau_{{\rm max}}$, follows the prediction for a one-dimensional
wire-geometry with disorder~\cite{Gopar2010,Pena2014}
\begin{equation}
   p\!\left(T\right)=C\frac{\sqrt{{\rm arccosh}\!\left(T^{-1/2}\right)}}{T^{3/2}\left(1-T\right)^{1/4}}{\rm exp}\!\left(-\frac{\xi^{\prime}}{2L}{\rm arccosh}^{2}\left(T^{-1/2}\right)\right),\label{eq:loc-dist}
\end{equation}
with $C$ being a normalization constant. The effective localization length $\xi^{\prime}=-2L/\left\langle {\rm
ln}T\right\rangle$ (the brackets here mean an average over different random realizations of the positions of the
hard-wall scatterers) can be determined from the numerical data. Figure \ref{fig:Distributions}(a),(b) shows the
comparison of the numerically calculated histograms of $\tau$ and $T$, respectively, and their analytical
predictions~\eqref{eq:chaos-dist} and~\eqref{eq:loc-dist}. We find that in both cases, the numerical data fits very well
the analytical formulae, which confirms our assumptions about the scattering dynamics being chaotic or localized for the
two different situations considered.

% Bibliography
%\bibliography{refs}

\end{document}